\documentclass[aoas,MSNbibl,nameyear,seceqn,dvips]{arximspdf}
\usepackage{graphicx}
\doi{10.1214/13-AOAS669} 
\volume{7}
\issue{4}
\pubyear{2013}
\firstpage{1960}
\lastpage{1982}

\makeatletter

\let\sv@tabnotetext\tabnotetext
\let\sv@tabnotemark@fmt\tabnotemark@fmt
\long\def\legend#1{{\let\tabnote@indent\leavevmode\sv@tabnotetext[]{}{#1}}}

\newcommand{\hdots}{\ldots}

\newcommand{\eps}{\varepsilon}
\newcommand{\Yv}{\mathbf{Y}}
\newcommand{\Xv}{\mathbf{X}}
\newcommand{\uv}{\mathbf{u}}
\newcommand{\Iv}{\mathbf{I}}
\newcommand{\Ov}{\mathbf{0}}

\newcommand{\Cv}{\mathbf{C}}
\newcommand{\Bv}{\mathbf{B}}
\newcommand{\Zv}{\mathbf{Z}}
\newcommand{\kapv}{{\bolds\kappa}}
\newcommand{\epsv}{\bolds\varepsilon}
\newcommand{\thetav}{{\bolds\theta}}
\newcommand{\expect}{\mathrm{E}}
\newcommand{\var}{\operatorname{Var}}

\def\d{\delta}
\setattribute{abstract}   {width}  {295pt}

\makeatother

\begin{document}
\begin{frontmatter}

\title{Quantifying intrinsic and extrinsic noise in gene transcription
using the linear noise approximation: An
application to single cell data\thanksref{T1}}
\runtitle{Quantifying intrinsic and extrinsic noise using the LNA}

\thankstext{T1}{The Centre for Cell Imaging has been supported
through BBSRC REI Grant BBE0129651.}

\begin{aug}
\author[A]{\fnms{B\"arbel}~\snm{Finkenst\"adt}\corref{}\thanksref{t00,m1}\ead[label=e1]{B.F.Finkenstadt@warwick.ac.uk}},
\author[B]{\fnms{Dan~J.}~\snm{Woodcock}\thanksref{t00,m1}\ead[label=e2]{D.J.Woodcock@warwick.ac.uk}},
\author[C]{\fnms{Michal}~\snm{Komorowski}\thanksref{tt0,m2}\ead[label=e3]{mkomor@ippt.gov.pl}},
\author[D]{\fnms{Claire~V.}~\snm{Harper}\thanksref{t0,m4}\ead[label=e4]{Claire.Harper@manchester.ac.uk}},
\author[E]{\fnms{Julian~R.~E.}~\snm{Davis}\thanksref{t11,m4}\ead[label=e5]{Julian.Davis@manchester.ac.uk}},\\
\author[D]{\fnms{Mike~R.~H.}~\snm{White}\thanksref{t11,m4}\ead[label=e6]{Mike.White@manchester.ac.uk}}
\and
\author[B]{\fnms{David A.} \snm{Rand}\thanksref{t00,t2,m1}\ead[label=e7]{D.A.Rand@warwick.ac.uk}\ead[label=u1,url]{http://www.foo.com}}
\runauthor{B. Finkenst\"adt et al.}
\affiliation{University of Warwick\thanksmark{m1}, Polish Academy of
Sciences\thanksmark{m2} and\break  University of Manchester\thanksmark{m4}}
\address[A]{B. Finkenst\"adt\\
Department of Statistics\\
University of Warwick\\
Coventry, CV47AL\\
United Kingdom\\
\printead{e1}}
\address[B]{D. J. Woodcock\\
D. A. Rand\\
Systems Biology Centre\\
University of Warwick\\
Coventry, CV47AL\\
United Kingdom\\
\printead{e2}\hspace*{5.8pt}\\
\hphantom{E-mail: }\printead*{e7}}
\address[C]{M. Komorowski\\
Polish Academy of Sciences\\
Division of Modelling in Biology\\
\quad and Medicine\\
02-106 Warszawa\\
Poland\\
\printead{e3}}
\address[D]{C. V. Harper\\
M. R. H. White\\
Faculty of Life Sciences\\
University of Manchester\\
Manchester, M13 9PT\\
United Kingdom\\
\printead{e4}\\
\hphantom{E-mail: }\printead*{e6}}
\address[E]{J. R. E. Davis\\
Faculty of Medical and Human Sciences\\
University of Manchester\\
Manchester, M13 9PT\\
United Kingdom\\
\printead{e5}}
\end{aug}

\thankstext{t00}{Supported by BBSRC and EPSRC (GR/S29256/01,
BB/F005814/1) and
EU BIOSIM Network Contract 005137.}

\thankstext{tt0}{Supported from University of Warwick, Department of Statistics,
and the Foundation for Polish Science
under the program Homing Plus HOMING 2011-3/4.}

\thankstext{t0}{Supported by The Professor John Glover Memorial Postdoctoral
Fellowship.}

\thankstext{t11}{Supported by a Wellcome Trust Programme Grant 67252.}

\thankstext{t2}{Supported by an EPSRC Senior Fellowship
(EP/C544587/1).}

\received{\smonth{1} \syear{2013}}
\revised{\smonth{6} \syear{2013}}

%
\begin{abstract}
A central challenge in computational modeling of dynamic biological
systems is parameter inference from experimental time course
measurements. However, one would not only like to infer kinetic
parameters but also study their variability from cell to cell. Here we
focus on the case where single-cell fluorescent protein imaging time
series data are available for a population of cells. Based on van
Kampen's linear noise approximation, we derive a dynamic state space
model for molecular populations which is then extended to a
hierarchical model. This model has potential to address the sources of
variability relevant to single-cell data, namely, intrinsic noise due
to the stochastic nature of the birth and death processes involved in
reactions and extrinsic noise arising from the cell-to-cell variation
of kinetic parameters. In order to infer such a model from
experimental data, one must also quantify the measurement process where
one has to allow for nonmeasurable molecular species as well as
measurement noise of unknown level and variance. The availability of
multiple single-cell time series data here provides a unique testbed
to fit such a model and quantify these different sources of variation
from experimental data.
\end{abstract}

%

\end{frontmatter}

\section{Introduction}
\label{sintro}
The effect of stochasticity (``\emph{noise}'') on linear and nonlinear
dynamical systems has
been studied for some time, yet significant new aspects continue to be
discovered. Here we consider
population dynamical systems, that is, systems which model the dynamics
of species
of populations stochastically by birth and death processes. This
modeling framework has been widely applied in many scientific fields,
including molecular biology, ecology, epidemiology and chemistry.
Examples include predator-prey population dynamics [\citet{McKane}],
SIR-type epidemic modeling [\citet{Simoes}], genetic networks [\citet
{Scott2006}],
molecular clocks [\citet{Gonze2002}] and biochemical networks [\citet
{Wilkinson}].

Basic cellular processes such as gene expression are
fundamentally stochastic with randomness in molecular machinery and
interactions leading
to cell-to-cell variations in mRNA and protein levels. This
stochasticity has
important consequences for cellular function and it is therefore
important to
quantify it [\citet{Thattai}, Paulsson (\citeyear{Paulsson2004,Paulsson2005}),
\citet{Swain2002}].
\citet{elowitz2002sge} defined \emph{extrinsic noise} in gene
expression in terms of fluctuations in the amount or activity of
molecules such as regulatory proteins and polymerases which in
turn cause corresponding fluctuations in the output of the gene.
They pointed out that such fluctuations represent sources of
extrinsic noise that are global to a single cell but vary from one cell
to another.
On the other hand, \emph{intrinsic noise} for a given gene was
defined in terms of the extent to which the activities of two
identical copies of that gene in the same intracellular environment
fail to correlate because of the random microscopic events that
govern the timing and order of reactions.
We can therefore consider that much of what makes up extrinsic noise
can be expressed mathematically in terms of the
stochastic variation between kinetic parameters
across a population of cells.

The availability of
replicate single cell data provides us with a unique
opportunity to estimate and explicitly quantify such between-cell variation.
Recent developments in fluorescent microscopy technology allow for
levels of reporter proteins such as green fluorescent protein (GFP) and
luciferase to be measured in vivo in
individual cells [\citet{Harper2011}]. Here, an important issue is to
relate the unobserved dynamics of expression of the gene under consideration
to the observed fluorescence levels of the reporter protein [\citet{Fink2008}].
This is facilitated by knowledge of the kinetic parameters associated
with the
translational and degradational processes of the reporter protein and mRNA.
In this study we present a methodology for estimating these
rates and their cell-to-cell variation. The approach can be seen
as an example of a general modeling framework which has the potential to
explicitly quantify and decouple both intrinsic and extrinsic noise
in population dynamical systems.

The structure of our study is as follows:
starting with a single cell stochastic model, we introduce the modeling
approach and inference methodology.
This is then
extended toward a population of
cells by introducing a Bayesian hierarchical structure where the
cell-to-cell variation in
certain parameters such as degradation rates is quantified
by a probability distribution. We also introduce the basic idea of the
linear noise approximation (LNA)
which is a key ingredient to rendering inference computationally efficient
given the complexity of a hierarchical model and the amount of
data. The performance of the Markov chain Monte Carlo
(MCMC) estimation algorithms is first tested on simulated data from the model
and results are presented for both simulated and real data.

\section{Model of gene expression and data}
\begin{figure}[b]

\includegraphics{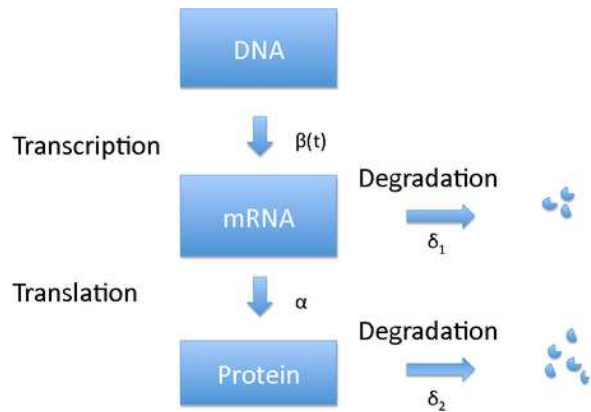}

\caption{The expression of a gene into its protein is determined by
four processes: transcription, translation, mRNA degradation and
protein degradation.}
\label{diagram}
\end{figure}

Our stochastic model for a single cell follows the general model of gene
expression [\citet{Paulsson2005}] shown in Figure~\ref{diagram}. The
active gene transcribes mRNA which is then translated into protein.
During this process both mRNA and protein are degraded.
%
We wish to infer this model from multiple single cell protein imaging
time series
of the sort shown in Figure~\ref{data1} resulting from two types of
experiments (see Appendix \ref{appexp} for technical details).
In the first experiment (Figure~\ref{data1}, left panel)
the synthesis of protein was inhibited by adding the translational
inhibitor Cycloheximide. In the second experiment
(Figure~\ref{data1}, right panel)
%
\begin{figure}

\includegraphics{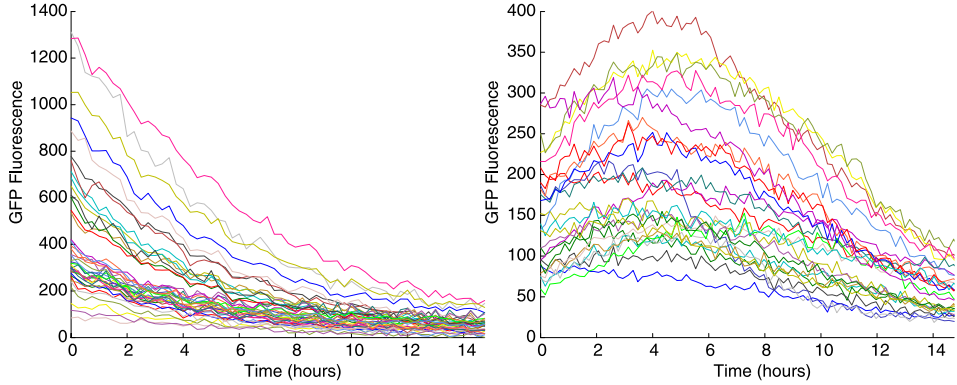}

\caption{\textup{Left}: observed fluorescence level from translation inhibition
experiment (40 cells, 59 observations per cell). \textup{Right}: observed
fluorescence level from transcription inhibition
experiment (25~cells, 88 observations per cell).
For both experiments measurements were taken simultaneously in cells
every 5 minutes.}
\label{data1}
\end{figure}
the transcriptional inhibitor Actinomycin D
was added to block the synthesis of mRNA.
In both cases only the fluorescent form of the protein is observable.
Single cell images of green fluorescent
protein (GFP) molecules were
collected
every 5 min at
discrete time points $t_i;i=1,\ldots,T$,
and quantified to give a time series $Y(t_i)$.
The images were collected simultaneously for a collection of
cells.\vadjust{\goodbreak}
Allowing for an
appropriately formulated measurement process,
our aim is to provide a statistical methodology for estimating
the kinetic parameters and for quantifying their variability between cells.
The model of gene expression in Figure~\ref{diagram} constitutes
a system of two different\vspace*{1pt} molecular subpopulations, namely, mRNA and
protein, with
state vector $X(t)=(X_1(t),X_2(t))^T$ where
%
\begin{table}[b]
\caption{Summary of reactions in the standard model of
gene expression}
\label{events}
\begin{tabular*}{\tablewidth}{@{\extracolsep{\fill}}lcc@{}}
\hline
\textbf{Event} & \textbf{Effect} & \textbf{Transition rate} \\
\hline
Transcription & $ (X_1,X_2) \to(X_1+1,X_2) $ & $ w_1= \beta(t) $ \\
Degradation of mRNA & $ (X_1,X_2) \to(X_1-1,X_2) $ & $ w_2= \delta_1
X_1(t) $ \\
Translation & $ (X_1,X_2) \to(X_1,X_2+1) $ & $ w_3= \alpha X_1(t) $ \\
Degradation of protein & $ (X_1,X_2) \to(X_1,X_2-1) $ & $ w_4= \delta
_2 X_2(t) $ \\
\hline
\end{tabular*}
\end{table}
$X_i(t), i=1,2$, denotes the number of molecules of each species,
respectively. There are
$m= 4$ possible reactions (transcription, degradation of the mRNA,
translation, degradation of the protein),
where a reaction of type $j$ changes $X(t)$ to $X(t)+v_j$
with
\[
v_1= \pmatrix{1
\cr
0},\qquad v_2= \pmatrix{-1
\cr
0},\qquad
v_3= \pmatrix{0
\cr
1},\qquad v_4= \pmatrix{0
\cr
-1},
\]
called stoichiometric vectors.
Each reaction occurs at a rate $w_j(X(t))$ as summarized in Table~\ref{events}.

\section{Inference for reaction networks}
Reaction networks such as the one introduced above constitute
continuous time Markov jump processes
and thus satisfy the Chapman--Kolmogorov equation\vadjust{\goodbreak}
for which one can obtain the forward form known as the master equation
(ME) describing the evolution of the probability $ P(X_1=n_1,
X_2=n_2;t) $.
Although an exact numerical simulation algorithm is provided [\citet
{Gillespie1977}], the
ME is rarely tractable and, hence, an explicit formula for the exact
likelihood is not available for parameter inference.
Additionally, longitudinal data from
such systems are usually discrete in time and only partially observed,
that is, not all molecular species are measurable during the experiment.
This poses further challenges to the estimation problem.

One way forward is to consider suitable approximations of the
likelihood function.
In particular, the diffusion approximation describing the process by a
set of It\^o stochastic
differential equations (SDEs), also called chemical Langevin equations,
has been of use in this context.
This approach is rigorously modeling the intrinsic noise of
the stochastic dynamics of the kinetic processes provided that the
assumptions of the SDE
approximation itself are valid. However, the associated likelihood is
also usually intractable and approximations have to be considered using
numerical simulations [Golightly and Wilkinson (\citeyear{Golightly2005,Golightly2008}),
\citet{Heron2007}].
Here, the basic idea, which has earlier been considered in econometric
applications of SDEs
to discretely observed data [\citet{Elerian2001,Durham2001}], is to
assume a Gaussian approximation to the transition density
[\citet{Kloeden1999}] for a sufficiently small time interval and to
augment the discretely observed data, along with any other unobserved
variables, by introducing a number of latent data points, thus creating
a fine virtual discrete time grid for
which the normal approximation is valid.
In particular, for partially observed systems the resulting estimation
algorithms are challenging to implement and computationally intensive
even for single time series because the dimension of the resulting
posterior density becomes very large.
\citet{Heron2007} and \citet{Golightly2008} consider additional
measurement error in a partially observed system
but assume that the observed data is measured at the correct population
level and that the variance of the measurement error is known. Attempts
to incorporate realistic assumptions about the measurement process have
been extremely limited within this framework and applications have only
been on artificially simulated
discretized time series data from chemical networks.

Another approach that has received attention in theoretical studies of
chemical networks
[\citet{Elf2003,Paulsson2004}]
is the linear noise approximation (LNA) [Van~Kampen
(\citeyear{VanKampen1976,VanKampen1997})].
This decomposes the system into a set of ordinary differential
equations (ODEs) for the mean
and a set of linear SDEs with Gaussian transition densities
for the fluctuations around the mean. While the validity of the LNA is
the subject of ongoing research [\citet{Wallace2012}],
its use for Bayesian inference
in chemical networks was first suggested and studied for some example
networks in \citet{Komic2009}
who also provide an application to single cell time series experimental data.
The LNA has the advantage that the approximated likelihood
is multivariate Gaussian. Hence, it is
straightforward to incorporate both Gaussian measurement error (with
unknown variance)
and a model relating the observed imaging data to the unobserved
molecular populations,
some of which may not be measurable.

\section{Linear noise approximation}
The LNA as formulated by Kurtz (\citeyear{Kurtz1971,Kurtz1981}) is derived
directly from the underlying Markov jump process and
is valid for any time interval of fixed length.
However, here we follow a simplified derivation of the LNA by \citet
{Wallace} [see also \citet{Wilkinson}] which
is adequate for our purposes. Readers should consult \citet{Kurtz1981}
for the precise assumptions and validity of the approximation.

Consider an approximation for a
fixed time interval of length $\tau$ where for each reaction $j$ we
define $K_j$ to be the number of
events of type $j$ that occur within the interval of length $\tau$.
$K_j$ depends on the rate $w_j$ and on $\tau$.
The system state vector can then be updated with the stoichiometric
vectors $v_j$ as
%
%
\begin{equation}
\label{update} X(t+\tau)=X(t)+ \sum_{j=1}^{m}
v_j K_j\bigl(w_j\bigl(X(t)\bigr),\tau
\bigr).
\end{equation}
Under the assumption that
$\tau$ is small enough so that the rate function $w_j(X(t)) $ can be
considered constant over $[t,t+\tau)$ for all $j$, known as the first
leap condition [\citet{Gillespie2001}],
the distribution of each $K_j$ is Poisson with mean and variance
$ \expect[K_j]=\var[K_j]=w_j(X(t))\tau$. If, furthermore, %
$\tau$ is large enough so that $w_j(X(t))\tau$ is large for all $j$
(second leap condition),
then $K_j$ is Gaussian
$K_j \sim N(w_j(X(t))\tau,w_j(X(t))\tau)$.
With the Gaussian assumption the updating rule in equation (\ref
{update}) becomes
%
%
\begin{equation}
\label{Lang} X(t+\tau)=X(t)+ \tau\sum_{j=1}^{m}
v_j w_j + \sqrt{\tau} \sum
_{j=1}^{m} v_j \sqrt{w_j}
\eps_j,
\end{equation}
where $\eps_j \sim N(0,1),j=1,\ldots,m$, are independent standard
normal random variables. For
simplicity of notation we use $w_j$ instead of $w_j(X(t))$.
The LNA makes the ansatz
%
%
\begin{equation}
\label{LNA1} X(t)= \Omega\phi(t) + \sqrt{\Omega} \zeta(t),
\end{equation}
that is, the process $X(t)$ can be written as the
deterministic solution of the macroscopic equations $\phi(t)=(\phi
_1(t), \phi_2(t))^T$
of the concentrations plus a residual stochastic process $\zeta
(t)=(\zeta_1(t), \zeta_2(t))^T$
where both components are scaled appropriately by the volume $\Omega$
of the system.
\citet{VanKampen1997} derives the LNA from a system size expansion of
the ME
where terms of first order give the macroscopic rate equations or ODEs,
and terms of second order give
a set of SDEs for the stochastic process $\zeta(t)$ (see Appendix \ref
{appLNA}).
A simplified derivation (see Appendix \ref{appLNA}) shows that the
LNA also provides a
first order Taylor approximation to equation (\ref{Lang}).
The macroscopic solutions for our model are
%
%
\begin{eqnarray}
\label{ode} \frac{d \phi_1 (t)}{d t} & = & \beta(t) - \delta_1
\phi_1(t),
\nonumber\\[-8pt]\\[-8pt]
\frac{d \phi_2 (t)}{d t} & = & \alpha\phi_1(t) - \delta_2 \phi
_2(t).
\nonumber
\end{eqnarray}
The residual stochastic process is characterized
by
%
%
\begin{equation}
\label{noise} d \zeta(t)= J \zeta(t) \,dt + B(t) \,d W(t),
\end{equation}
where $W=(W_1,W_2)^T$, $W_1,W_2$ are independent Wiener processes,
\[
B(t)= \pmatrix{\sqrt{\beta(t) + \delta_1 \phi_1(t) } &
0
\cr
0 & \sqrt{ \alpha\phi_1(t) + \delta_2
\phi_2(t) }}
\]
and
\[
J= \pmatrix{-\delta_1 & 0
\cr
\alpha& -\delta_2}.
\]
Due to the linearity of (\ref{noise}), the transition densities
$P(\zeta(t_{i+1})|\zeta(t_i))$
for arbitrary time length are Gaussian. For our model $J$ is
time-independent and, hence,
the mean is given by
\[
\mu(t_{i})=e^{J\Delta_i}\zeta(t_i),
\]
that is, the solution for the deterministic part of (\ref{noise}) from
a starting value $\zeta(t_i)$,
where $\Delta_i=t_{i+1}-t_i$ is the length of the interval. The
covariance matrix is
%
%
\begin{equation}
\label{sigma} \Sigma(t_{i})=\int_{t_i}^{t_{i+1}}e^{J(t_{i+1}-s)}B(s)B(s)^T
\bigl(e^{J(t_{i+1}-s)}\bigr)^T \,ds.
\end{equation}
The approach does not require equidistant measurements.
In general, the integrals needed for the mean and covariance of the
transition densities
arising from the LNA can be either determined explicitly or computed
numerically.
Since $P(\zeta(t_{i+1})|\zeta(t_i))$ is $N(\mu(t_{i}),\Sigma(t_{i}))$,
we have that $ P(X(t_{i+1})|X(t_i))$ is $N( \Omega\phi(t) + \sqrt {\Omega} \mu(t_{i}),\Omega\Sigma(t_{i}))$.
Thus, the LNA estimates the variances of the species abundances and the
covariances between them and
a transition from $X(t_i)$ to $X(t_{i+1})$ follows the state space equation
%
%
\begin{equation}
\label{bigeuler} X(t_{i+1}) = F(t_i) X (t_i)
+c(t_i)+ \eps_{t_{i}},\qquad \eps _{t_{i}} \sim N\bigl( \Ov,
\Sigma_{\eps}(t_{i}) \bigr),
\end{equation}
where for our model $F(t_i)=e^{J\Delta_i}$, $ c(t_i)=\Omega [
\phi(t_{i+1}) - e^{J\Delta_i} \phi(t_{i})  ] $ and $\Sigma
_{\eps}(t_{i})=\Omega\Sigma(t_{i})$.

In general, we assume a measurement equation of the form
%
%
\begin{equation}
\label{meas} Y(t_i)= \kappa X(t_i) +
u(t_i),
\end{equation}
where $\kappa$ is a matrix,
$Y(t_i)$, $ i=1,\ldots,T$, are the observed data for a single cell,
and the $u(t_i) \sim N( \Ov, \Sigma_u(t_{i}))$ represent
measurement errors with covariance matrix $\Sigma_u(t_{i})$.
Note that $\kappa$ accounts for the assumption that
imaging data may be proportional to molecular population size.
State variables that cannot be measured reduce the rank of the matrix.
As in all our applications below, only the protein is imaged,
$\kappa$ will be a scalar and $u(t_i)$ will be one-dimensional.

Let $\Xv=(X(t_1),\ldots,X(t_T))^T$ and $\Yv=(Y(t_1),\ldots,Y(t_T))^T$.
The joint density of $\Yv$ or likelihood $L(\Yv|\theta)$ can be
obtained by writing the system of
equations (\ref{bigeuler}) for all discrete time points $t_i$, $
i=1,\ldots,T$ as
%
%
\begin{equation}
\Bv\Xv= \Zv+ \Cv+ \epsv,
\end{equation}
where
\begin{eqnarray*}
\Bv&=& \pmatrix{ \Iv& \Ov& \cdots&\cdots& \cdots& \Ov
\cr
- e^{J\Delta_1} & \Iv&
\ddots& & & \vdots
\cr
\Ov& \ddots&\ddots&\ddots& & \vdots
\cr
\vdots& \ddots&
\ddots& \ddots& \ddots& \vdots
\cr
\vdots& & \ddots& \ddots& \ddots& \Ov
\cr
\Ov&
\cdots& \cdots& \Ov& - e^{J\Delta_{T-1}} & \Iv},
\\
\Cv&=&\bigl(c(t_0),\ldots, c(t_{T-1})\bigr)^T,\qquad \Zv= \bigl(
e^{-J\Delta_0} X(0), \Ov\hdots\Ov \bigr)^T
\end{eqnarray*}
and $\epsv\sim N(\Ov,\Sigma_{\epsv})$ with
covariance matrix $ \Sigma_{\epsv}= \operatorname{diag}( \Sigma_{\eps
}(t_0),\ldots, \Sigma_{\eps}(t_{T-1}) )$.

It follows that
%
%
\begin{equation}
\label{distx} \Xv\sim N\bigl(\Bv^{-1}(\Cv+ \Zv),\Bv^{-1}
\bigl(\Sigma_{\epsv} + \var (\Zv)\bigr) \bigl(\Bv^{-1}
\bigr)^T\bigr).
\end{equation}
The matrix formulation of the observational equation (\ref{meas}) for
all time points is
%
%
\begin{equation}
\label{fullmeas} \Yv=\kapv\Xv+ \uv,
\end{equation}
where $\kapv$ is the matrix that has $\kappa$ along the diagonal and is
zero otherwise, and $\uv=(u(t_1),\ldots,u(t_T))^T$ with $\uv\sim
N(\Ov,\Sigma_{\uv})$. Hence, the likelihood $L(\Yv|\theta)$ is [see
also \citet{Komic2009}]
%
%
\begin{equation}
\label{lik1cell}\quad  \Yv\sim N\bigl(\kapv\Bv^{-1}(\Cv+ \Zv),
\kapv\Bv^{-1}\bigl(\Sigma_{\epsv
} + \var(\Zv)\bigr) \bigl(
\Bv^{-1}\bigr)^T \kapv^T + \Sigma_{\uv}
\bigr).
\end{equation}
We assume that measurement errors $u(t_i)$ are i.i.d. normal
so that $\Sigma_{\uv}=\sigma^2_u \Iv$, but the approach can be
adapted to other specifications of the error covariance matrix.

In situations where the sample size and/or the dimension of $\Yv$ is
large, the density corresponding to (\ref{lik1cell}) may be
expensive to compute, as it requires inversion of large matrices. Since
the model for $Y(t_i)$ can be written in state space form defined by
equations (\ref{bigeuler}) and (\ref{meas}), an equivalent form of
(\ref{lik1cell}) that is easier to handle is obtained
using the prediction error decomposition with log likelihood
%
%
\begin{eqnarray}
\label{PED} \log L(\Yv; \theta) & = &\sum_{i=1}^T
\log f(Y_{t_i} | Y_{t_1},\ldots,Y_{t_{i-1}}; \theta)
\nonumber\\[-8pt]\\[-8pt]
& = & \sum_{i=1}^T \biggl[-
\frac{\operatorname{dim}(Y_{t_i})}{2} \log2 \pi - \frac{1}{2} \log |R_{t_i}| -
\frac{1}{2} e_{t_i}^T R_{t_i}^{-1}e_{t_i}
\biggr],\nonumber
\end{eqnarray}
where $e_{t_i}=Y_{t_i} - \hat{Y}_{t_i|t_{i-1}}$ is the prediction error,
$ \hat{Y}_{t_i|t_{i-1}}=E(Y_{t_i} | Y_{t_1},\ldots,\break Y_{t_{i-1}};
\theta)$ is the optimal predictor of $Y_{t_i}$ given the information
up to time
$t_{i-1}$ and $R_{t_i}$ is the variance matrix of the prediction error.
These quantities can be computed as part of the
Kalman filter recursions (see Appendix \ref{appKL}).

%
\section{A hierarchical model for multiple cells}
The full data matrix of an experiment
contains $N$ multiple imaging time series
\[
\tilde{ \Yv}=\bigl(\Yv^{(1)},\Yv^{(2)},\ldots,
\Yv^{(N)}\bigr),
\]
where $\Yv^{(i)}$ are the imaging data for a cell now indexed by $i;
i=1,\ldots,N$.
Bayesian hierarchical modeling [\citet{Gamerman}] provides a natural
framework to account for
the cell-to-cell variability of kinetic parameters.
Assuming that replicates are independent, the full likelihood for all
cells in the experiment is
%
%
\begin{equation}
\label{fullik} L(\tilde{ \Yv};\thetav)=\prod_{i=1}^{N}L
\bigl(\Yv^{(i)}|\theta^{(i)}\bigr),
\end{equation}
where $\theta^{(i)}$ denotes the vector containing all parameters for
cell $i$
and\break  $L(\Yv^{(i)}|\theta^{(i)})$ is the likelihood for a single cell
as derived above.
In contrast to assuming that a reaction $j$ in all cells is described
by exactly the same value of
the associated kinetic parameter $\theta_j$, in a hierarchical model
it is
a sample from a population distribution $p(\theta_j|\Theta_j)$ which
is governed by
an unknown parameter vector $\Theta_j$ quantifying the mean and variance
of each hierarchical parameter across the population of cells.
Let $\thetav=(\theta^{(1)},\theta^{(2)},\ldots,\theta^{(N)})$
denote the matrix of parameter vectors and
let $p(\thetav|\Theta)$ denote the joint distribution of $\thetav$ assuming
\[
p(\thetav|\Theta)=\prod_j p(
\theta_j|\Theta_j),
\]
where $\Theta$ is the vector of all hyperparameters.
In the hierarchical model we wish to infer upon the posterior $p(\Theta
|\tilde{\Yv})$,
%
%
\begin{equation}
\label{Bayes} p(\Theta|\tilde{ \Yv}) \propto L(\tilde{ \Yv};\thetav) p(\thetav |
\Theta) p(\Theta),
\end{equation}
where $p(\Theta)$ denotes the prior distribution of the
hyperparameters. This is achieved by formulating an
appropriate MCMC algorithm that samples from $p(\Theta|\tilde{ \Yv})$.
%
\subsection{Translation inhibitor experiment}
We start with the translation inhibition experiment, as it allows us to estimate
the protein half life which is then used as prior information for the
other experiment.
We assume that under the influence of the translational inhibitor the
level of protein synthesis drops down to
zero or possibly a small basal level $\tau_2$ in case inhibition is
not fully achieved while the initial protein
degrades at per capita rate $\delta_2$ [\citet{gordon2007}]. The
resulting model thus
does not depend on the mRNA process and simplifies to the
univariate case where $X(t)=X_2(t)$ with a single macroscopic equation
\[
\frac{d \phi_2 (t)}{d t} = \tau_2 - \delta_2
\phi_2(t).
\]
The LNA noise process is thus one-dimensional $\eta=(\eta_2)$, where
\[
B(t)= \sqrt{ \tau_2 + \delta_2 \phi_2(t)}
\]
and $J=-\delta_2$. The following parameters are assumed to be
hierarchical $\delta_2^{(i)}, \tau_2^{(i)}$, $\sigma_u^{(i)},
i=1,\ldots,N$. We reparameterize
$\tilde{\tau}_{2}^{(i)}=\kappa\tau_{2}^{(i)}$, which significantly
improves convergence of the estimation algorithm. Details on the
specification of the distributions of the parameters are given in
Appendix \ref{apptransla}. As the initial conditions $\phi_2^{(i)}(0)$
may be very different across cells, in particular, in experiments where
the cell behavior is not synchronized, we estimate them independently
for each cell rather than assuming a hierarchical structure. For
simplicity, the scaling parameter $\kappa$ is assumed to be
constant for all cells. %

\subsection{Transcription inhibitor experiment}
Similarly to the previous experiment, we assume that under the
influence of a transcriptional inhibitor
mRNA synthesis drops to some small basal level $\tau_1$ while the
initial amount of mRNA degrades and is also translated into protein
which then degrades.
The model is thus given by the full two-species model with $\beta
(t)=\tau_1$.
As only the protein is imaged, the measurement equation is formulated
in the same way as for the previous experiment.
We specify $\tau_1^{(i)}$, $\delta_1^{(i)}$, $\alpha^{(i)}$ and
$\sigma_u^{(i)}$
as hierarchical parameters (see Appendix \ref{apptranscript} for
details) and use the estimation results for the protein degradation parameter
from the translation inhibitor experiment as the informative prior. Our
simulation studies show that this is a crucial step,
as without this prior the other parameters were not identifiable due to
the mRNA not being observable.

%
\begin{table}
\tabcolsep=0pt
\caption{Results for simulated data}\label{simresults}
\begin{tabular*}{\tablewidth}{@{\extracolsep{\fill}}lcccc@{}}
\hline
\textbf{Parameter} & \textbf{True mean} & \textbf{Estimated mean} &
\textbf{True variance} & \textbf{Estimated variance} \\
\hline\\[-8pt]
\multicolumn{5}{@{}c@{}}{Translation inhibitor experiment} \\
[6pt]
$\tau_{2}$ & $3.675\cdot10^4$ & 182 $(54,1408)$ & $6.345\cdot10^8$ &
$25\cdot10^3$
$(2\cdot10^3,1.5\cdot10^6)$ \\[2pt]
$\tilde{\tau}_{2}$ & 3.675 & 3.61 $(2.76,4.55)$ & 6.345 & 9.48
$(4.84,18.70)$ \\[2pt]
$\delta_2$ & 0.576 & 0.56 $(0.54,0.57)$ & 0.005 & 0.004 $(0.002,0.006)$\\[2pt]
$\sigma_u^{2}$ & 12 & 11.86 $(11.02,12.54)$ & 3 & 3.89 $(1.34,7.06)$\\[2pt]
$\kappa$ & $10^{-4}$ & 0.02 $(0.00,0.05)$ & -- & -- \\
[6pt]
$\tau_{2}$ & 3.675 & 3.40 $(2.23,4.69)$ & 6.345 & 6.54 $(1.57,17.14)$ \\[2pt]
$\tilde{\tau}_{2}$ & 3.675 & 3.43 $(2.42,4.49)$ & 6.345 & 6.67
$(1.82,17.93)$ \\[2pt]
$\delta_2$ & 0.576 & 0.56 $(0.53,0.58)$ & 0.005 & 0.004 $(0.001,0.009)$\\[2pt]
$\sigma_u^{2}$ & 12 & 12.27 $(11.05,13.25)$ & 3 & 5.12 $(1.61,11.31)$\\[2pt]
$\kappa$ & 1 & 1.01 $(0.82,1.17)$ & -- & -- \\
[6pt]
\multicolumn{5}{@{}c@{}}{Transcription inhibitor experiment} \\
[6pt]
$\tau_{1}$ & 40 & 37.40 $(30.56,43.39)$ & 2 & 5.172 $(1.576,9.84)$ \\[2pt]
$\delta_1$ & 0.2 & 0.193 $(0.183,0.208)$ & 0.005 & 0.0080
$(0.00013,0.0173)$\\[2pt]
$\alpha$ & 3.5 & 3.731 $(2.557,4.940)$ & 2 & 1.483 $(0.684,4.056)$\\[2pt]
$\sigma_u^{2}$ & 10 & 9.124 $(8.275,10.144)$ & 2 & 1.615 $(0.363,4.602)$\\[2pt]
$\kappa$ & 0.25 & 0.239 $(0.221,0.254)$ & -- & -- \\
\hline
\end{tabular*}
\legend{True values and posterior estimates of the mean and
variance of the distribution of hierarchical parameters
for simulated data. $\kappa$ is not hierarchical. For\vspace*{1pt} the translation
inhibitor model two
cases are considered: large number of molecules with $\phi
_2^{(i)}(0)=2 \times10^6, \kappa=10^{-4}$ (top), and small
number of molecules with $\phi_2^{(i)}(0)=500, \kappa=1$ (bottom).
For the transcription inhibition model,
one case is considered with $\phi_1^{(i)}(0)=500, \phi
_2^{(i)}(0)=2000$. Estimates are medians (with 95\% interval in
brackets) of the posterior chains from 40 K iterations after
convergence is achieved. All rates are per hour.
The choice of rate parameters was
motivated by values that generate artificial data
with approximately similar dynamics as the real data and using
preliminary results from
fitting ODEs to aggregate data.}
\end{table}

%
\section{Results}

\subsection{Simulation studies}
In order to develop the MCMC estimation algorithms,
we generated artificial data of similar sample size and sampling
frequency as
the observed data for chosen parameter
values as displayed in Table~\ref{simresults}. Such simulation studies
are vital to developing the estimation algorithm,
assessing its performance, checking for bias and gaining an
understanding of the precision with which parameters can be estimated.
We also study the simpler case of the translation inhibition experiment
to compare estimation for two scenarios,
namely, a~system with a large and a small number of molecules where
the set parameter $\kappa$ was adjusted to give values in a similar
range so that the measurement error had a similar impact
in both scenarios. Artificial data was generated with exact intrinsic
stochasticity using the
stochastic simulation algorithm [\citet{Gillespie1977}] and normal
measurement error.

Table~\ref{simresults} summarizes estimation results for the
simulation study confirming that
estimation based on the LNA
is successfully reproducing posterior estimates with reasonable precision.
Retrieving both $\kappa$ and measurement error variance is a
significant achievement, not least because
it gives us an idea of the size of the molecular populations. The
simulation study shows that $\kappa$ is estimated with more
precision for the smaller molecular system. A possible reason for this
is that since the intrinsic noise
scales with factor $\sqrt{\kappa}$ between the measurement and
population level,
the information about the intrinsic noise is essential in identifying
$\kappa$.
For larger population sizes the trajectories become smoother and the
intrinsic noise will be
less informative about $\kappa$.
We thus conjecture that while it is the stochastic approximation which
facilitates calibration
of the model in terms of population size, it will be less successful in
doing so if the molecular population is large
and a simpler ODE approximation may be assumed to be adequate.

%
\begin{table}
\tablewidth=265pt
\caption{Results for experimental data}\label{dataresults}
\begin{tabular*}{\tablewidth}{@{\extracolsep{\fill}}lcc@{}}
\hline
\textbf{Parameter} & \textbf{Estimated mean} & \textbf{Estimated variance} \\
\hline\\[-8pt]
\multicolumn{3}{@{}c@{}}{Translation inhibitor experiment} \\
[6pt]
$\tau_{2}$ & 50.67 $(24.12,114.40)$ & 1086.40 $(206.41,6155.13)$ \\[2pt]
$\tilde{\tau}_{2}$ & 3.51 $(2.84,4.22)$ & 5.07 $(2.35,9.59)$ \\[2pt]
$\delta_2$ & 0.57 $(0.54,0.59)$ & 0.004 $(0.002,0.007)$\\[2pt]
$\sigma_u^{2}$ & 6.36 $(5.11,7.65)$ & 20.07 $(10.25,35.47)$\\[2pt]
$\kappa$ & 0.07 $(0.02,0.12)$ & -- \\
[6pt]
\multicolumn{3}{@{}c@{}}{Transcription inhibitor experiment} \\
[6pt]
$\tau_{1}$ & 3.53 $(2.64,5.02)$ & 9.38 $(3.79,23.90)$ \\[2pt]
$\tilde{\tau}_1$ & 1.53 $(1.10,2.55)$ & 4.29 $(0.93,21.12)$ \\[2pt]
$\delta_1$ & 0.13 $(0.12,0.15)$ & 0.004 $(0.001,0.013)$\\[2pt]
$\alpha$ & 3.93 $(3.27,4.80)$ & 6.09 $(3.35,11.11)$\\[2pt]
$\tilde{\alpha}$ & 0.44 $(0.34,0.62)$ & 0.08 $(0.04,0.17)$ \\[2pt]
$\sigma_u^{2}$ & 5.14 $(4.82,5.90)$ & 1.09 $(0.55,5.94)$\\[2pt]
$\kappa$ & 0.11 $(0.09,0.14)$ & -- \\
\hline
\end{tabular*}
\legend{Posterior estimates of the mean and variance of the
distribution of hierarchical parameters
for experimental data from translation and transcription inhibitor
experiments. $\kappa$ is not hierarchical. All rates are per hour.
Estimates were computed from posterior chains as described in
Table~\ref{simresults}.}
\end{table}

We also developed and studied performance of our estimation algorithm
for the transcription inhibitor experiment
via a simulation study as reported in Table~\ref{simresults}.
Inference is more challenging for the two-dimensional model
when only one variable is measurable. The parameter traces tend to be
more correlated
and more time is needed on fine-tuning the algorithm.
To enhance efficiency over the conventional Metropolis--Hastings algorithm,
we implemented a modified MCMC algorithm based on the
Metropolis--Hastings method.
In particular, we used block sampling [\citet{Gamerman}] in combination with
the multiple-try Metropolis (MTM) algorithm with antithetic sampling as
in \citet{craiu}.
The original MTM algorithm [\citet{LLW}] generates a number of
proposals for each parameter
and selects one with a probability that is proportional to its
likelihood. We then construct a backward step from the chosen proposal
so that the detailed balance condition is satisfied and then accept or
reject this proposal in the conventional Metropolis manner. The
antithetic multiple correlated try Metroplis (MCTM) method incorporates
negative correlation into this framework to maximize the Euclidean
distance between these proposals. Together with the reparameterization
it was found to
improve convergence of the estimation algorithm.
%

\subsection{Results for experimental data}
\begin{figure}

\includegraphics{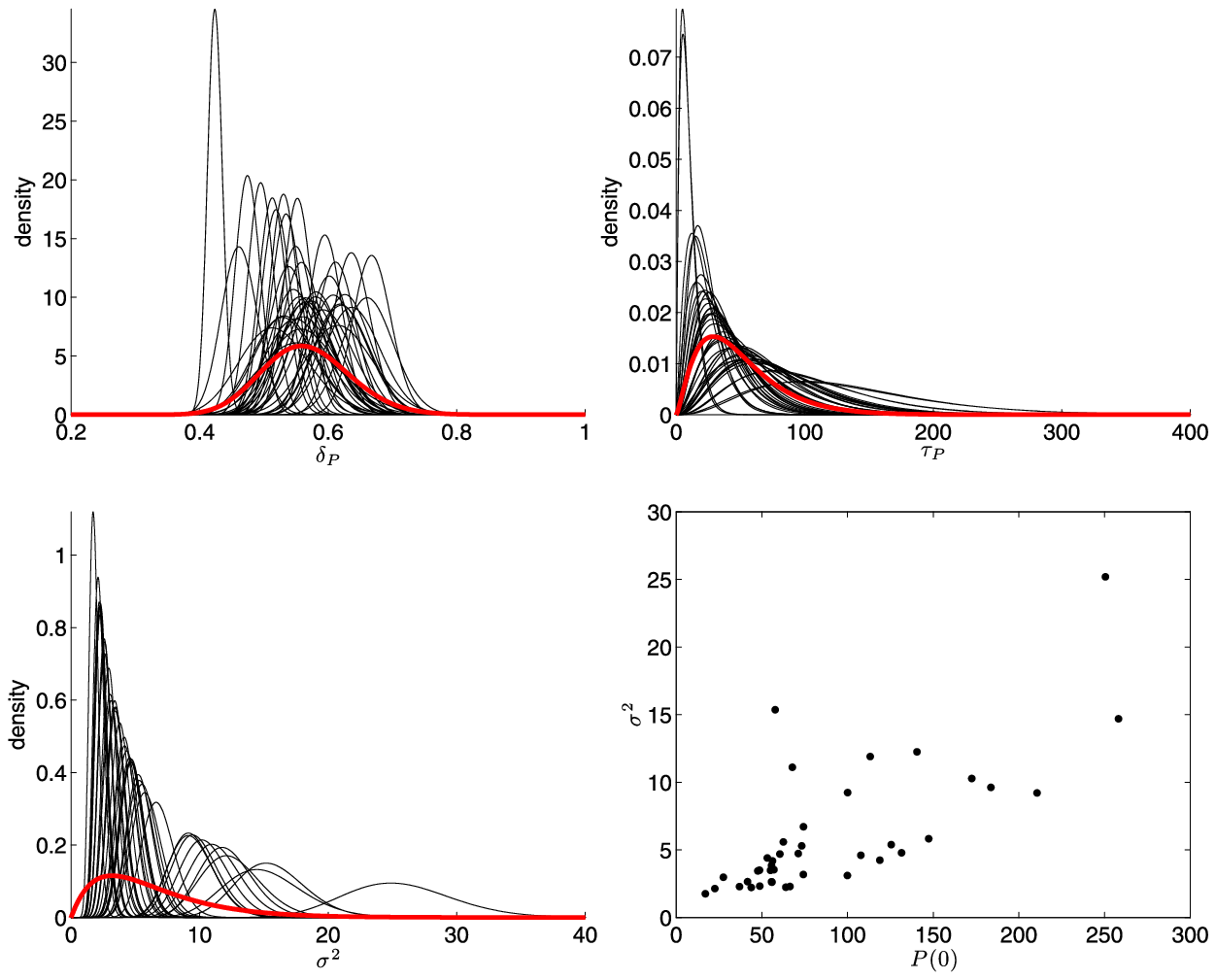}

\caption{\textup{Translation inhibitor experiment.} Estimated posterior
densities of parameters $\delta_2$ (top left), $\tau_2$ (top right)
and $\sigma^2_u$ (bottom left) from experimental data of the
translation inhibitor experiment. Black solid lines give estimated
posterior densities for the single cells using kernel estimation. Red
solid curve gives their estimated joint density as specified by the
hierarchical model. The scatterplot (bottom right) gives the estimated
standard deviation $\sigma_u$ of the measurement error against
estimated initial condition of the macroscopic solution $\phi_2(0)$
for each cell. The empirical Spearman correlation coefficient is 0.77.}
\label{fig3}
\end{figure}

The results for the real data are given in
Table~\ref{dataresults} and are plotted in Figures~\ref{fig3} and
%
\ref{fig4} for the
translation and transcription inhibition, respectively. Diagnostic
tests applied to the standardized prediction error
(\ref{KFerror}) computed for the mean posterior parameter estimates
indicate that residuals do not exhibit significant autocorrelation
and their distributions are compatible with normality. We use the
coefficient of variation CV (ratio of standard deviation to mean)
to compare between-cell variability of different parameters.

\subsubsection*{Degradation rates}
Out of all estimated rates it seems the degradation rates for protein
and mRNA exhibit the least
cell-to-cell variation.
The mean protein degradation rate was estimated to be around 0.576,
which corresponds to a half-life of
approximately 1.2 h. The estimated cell-to-cell variation in the
degradation rate is 0.063 (standard deviation) and
%
\begin{figure}[t!]

\includegraphics{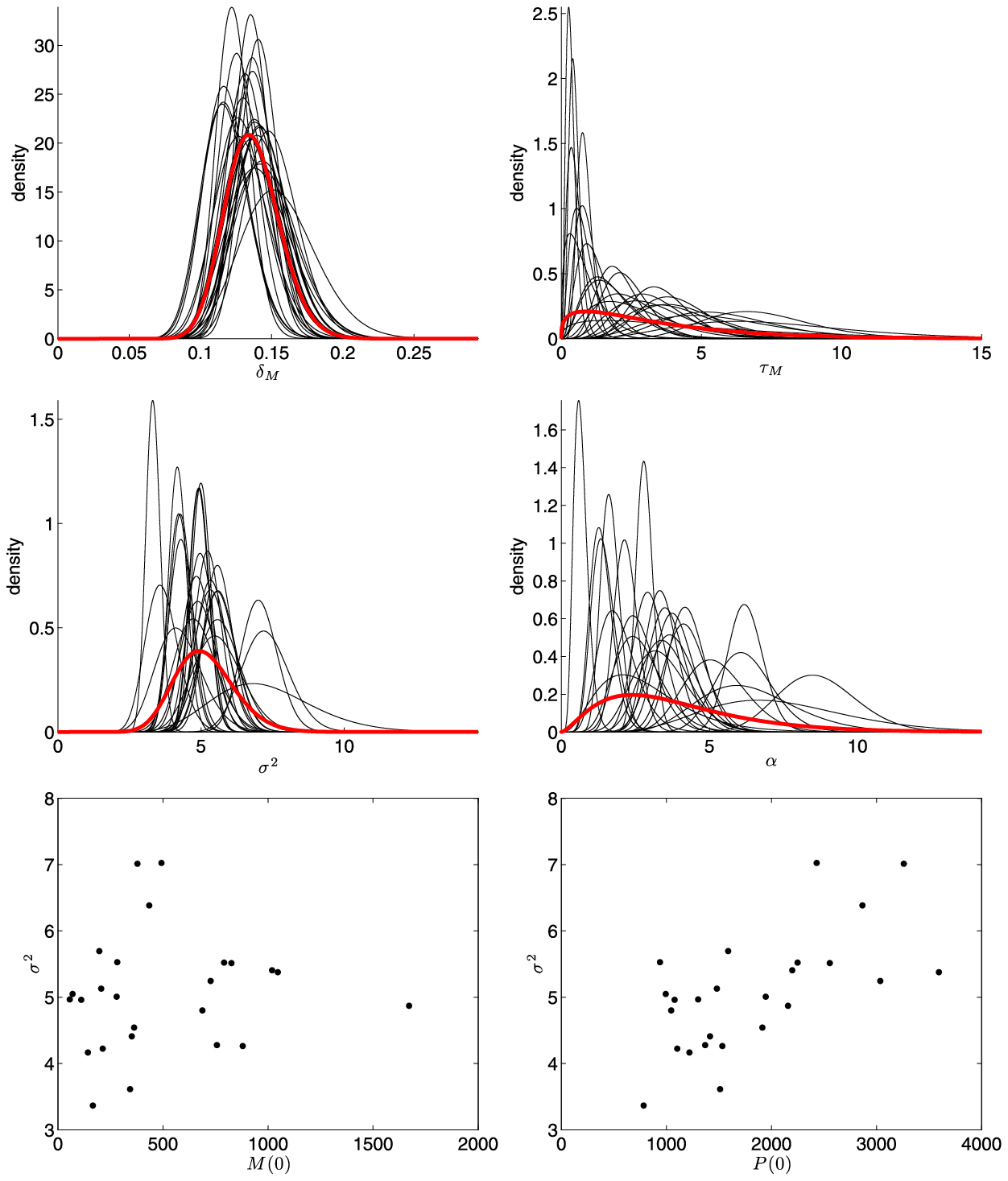}

\caption{\textup{Transcription inhibitor experiment.} Estimated posterior
densities of parameters $\delta_1$ (top left), $\tau_1$ (top right),
$\sigma^2_u$ (middle left) and $\alpha$ (middle right) from
experimental data of the transcription inhibitor experiment. Black
solid lines give estimated posterior densities for the single cells
using kernel estimation. Red solid curve gives their estimated joint
density as specified by the hierarchical model. The scatterplots give
the estimated standard deviation $\sigma_u$ of the measurement error
against estimated initial conditions of the macroscopic equations $\phi
_1(0)$ (bottom left) and $\phi_2(0)$ (bottom right). The empirical
Spearman correlation coefficients are 0.11 and 0.63, respectively.}
\label{fig4}
\end{figure}
the CV is 0.1, indicating that there is only small variation between
cells. The 2-sigma band for the estimated density
of the protein degradation rate of the hierarchical model (red line in
Figure~\ref{fig3} top left) is $(0.45\mbox{--}0.70)$. Thus, almost all cells in
the sample exhibit a mean protein half-life
between 1 h and 1.5 h. The mRNA degradation rate was estimated around
0.137 (Figure~\ref{fig4} top left), corresponding to a
half-life of 5 h. The cell-to-cell variation is about 0.02 (standard
deviation) and is also very small with a CV of 0.05. From the
2-sigma band we find that almost all cells have an mRNA half-life
between 4 h and 7 h.
\subsubsection*{Transcription and translation rates}
The rate $\tau_{1}$ under transcriptional inhibition is smaller than
the rate $\tau_{2}$ under translational inhibition.
This is reasonable, as population size for protein can be expected to
be larger than for mRNA. The CVs are 0.67 and 0.87 for $\tau_2$
and $\tau_1$, respectively, indicating considerably more between-cell
variability than for the degradation rates.
This may also be due to the fact that $\tau_1$ and $\tau_2$ will vary
with cell size and abundance.
The translation rate $\alpha$
is estimated around 4 with similar cell-to-cell variation as
transcription rates with a CV of 0.62.
We note that neither $\tau_{2}$ nor $\tau_{1}$ appear to be close to
zero, indicating
that neither treatment achieved complete inhibition.
Our estimation shows that degradation rates show less variability than
transcription rates.
It is interesting to speculate that this is because of the extra
variability that would arise from a
combination of the bursting structure that has been observed in genes
[\citet{Harper2011,Suter}] and the related effect of
chromatin remodeling and transcription factor variability.

\subsubsection*{Scaling factor $\kappa$, molecular population size
and measurement error}
The scaling factor $\kappa$ was estimated to be around 0.07 and 0.10
in the translation and
transcription inhibitor experiment, respectively. Both values overlap
largely in their posterior distribution, indicating
that the scaling is similar between the two experiments, which is
reasonable as the
experimental protocol with respect to taking the images was similar.
Simple plug-in estimates of the initial molecular population sizes can
be obtained
from the Markov chain traces.
For the translation inhibitor experiment the average (over the cells) initial
protein abundance is estimated to be around
1260 (median) with 95 \% central range (660--3660)
and similar for the transcription inhibitor experiment [1530 (median)
with 95 \% central range (880--3290)].
The average initial mRNA abundance
is smaller [370 (median) with 95\% central range (70--1150)].
There is considerable variation
between cells in the initial conditions and it becomes clear that the
variation seen in the data of Figure~\ref{data1} is predominantly due to the cell-to-cell variation in
initial population size.
The estimated variance $\sigma_{u}^2$ of the measurement error is of
similar size in both
experiments. We find that the individual cell estimates of the
measurement error variance correlate strongly
with the initial protein abundance of the cell in both experiments
(Figures~\ref{fig3} and
\ref{fig4}, bottom right). The correlation
between the measurement error variance and the initial mRNA population
(Figure~\ref{fig4}, bottom left)
is weaker, which is also reasonable as the imaging data are more
directly related to protein rather than mRNA abundance.

\section{Summary and discussion}
In this study we have introduced a Bayesian hierarchical model based on
the LNA that can be used in the context of stochastic compartmental
population models.
The model has the potential to tease out different sources of
variability that are relevant to inference about
transcriptional, translational and degradational processes at the
molecular level, namely:
\begin{itemize}
\item\textit{intrinsic stochasticity} due to the natural stochastic
nature of the birth and death processes involved in chemical reactions,
\item\textit{extrinsic variability}, that is, arising from the cell-to-cell
variation of kinetic parameters associated with these processes, and
\item\textit{measurement noise} which is additive and not part of the
dynamic process.
\end{itemize}
We focus on drawing inference about these sources of stochasticity from
real experimental time series data in molecular systems biology.
The two experiments considered here are an example of a scenario where
the use of a rich stochastic model in combination with a Bayesian
approach to inference can
deal with shortcomings such as unobservable population species.
It demonstrates that, provided the data are indeed of the kind to
display intrinsic stochasticity, that is, come from a single cell,
assuming a stochastic model is more informative than an ODE approach
even if the latter fits the mean well. Here it allowed us to decouple
the different sources of noise and to obtain an estimate of the scaling
factor $\kappa$, leading to inference about the size of the underlying
molecular species.
Pilot simulation studies are an essential tool not only to develop the
estimation algorithm but also to study parameter identifiability. The
posterior information about the kinetic parameters obtained here will
allow us further to study nonlinear transcription functions and spatial
characteristics of the gene under consideration, that is, Prolactin, in
spatio-temporal experiments using the same reporter construct.

Our simulation studies confirm that the LNA works well for inference in
our model, also for smaller molecular population sizes. This is in line
with results in
\citet{Komic2009}. Recently, \citet{Stathopoulos2013} combined the LNA
with their Riemann manifold MCMC sampling methods and found that this
combined approach is both statistically and computationally efficient
for reaction networks also in the case of smaller populations. A study
by \citet{Wallace2012} provides new insight on the range of validity of
the LNA and supports its applicability to systems such as studied in
this paper.
The rigorous presentation of the LNA by Kurtz (\citeyear{Kurtz1971,Kurtz1981})
clearly states the underlying assumptions, in
particular, the need to restrict to a finite time horizon before taking
the limit $\Omega\to\infty$ except in the case when the ODE has a
single stable stationary solution. Therefore, this method cannot be
expected to work when molecular numbers are too small. The threshold
population size above which the method gives good results will depend
on the nature of the dynamical system and the time horizon set. For
example, a system near to a bifurcation or with multiple attractors is
expected to have a significantly larger threshold than one with a
single globally attracting attractor.
Similarly, the LNA is likely to struggle when the ODE solution in the
LNA is not equal to the mean value of the stochastic process.
An obvious remedy is to reset the LNA by allowing for the initial
conditions of the ODE to evolve conditional on the observed data and
recent work
by \citet{Fearnhead2013} shows that this improves estimation in a
nonlinear predator-prey interaction model. We did not encounter this
problem because in our model the Jacobian does not depend on the
deterministic solution. We are also currently studying the use of the
LNA for inference in systems with small populations and with nonlinear
switch-type transcription functions and so far have found that the LNA
copes well with the nonlinearity of the switch function and only begins
to break down for extremely small population sizes.

The hierarchical model provides a useful and natural approach to study
and quantify the cell-to-cell variability between kinetic parameters.
The resulting model is complex and the use of the LNA has been crucial
to facilitate its inference. Computational feasibility and ability to
link between the model and experimental data through a realistically
modeled measurement process are essential for studying problems in
systems biology where in the future many more single cell data sets
from complex systems will become available.

%
\begin{appendix}\label{app}
\section*{Appendix}
\subsection{Experiment}\label{appexp}
GH3-DP1 cells containing a stably integrated 5kb human
prolactin-destabilised EGFP reporter gene [\citet{Harper2011}]
were grown in Dulbecco's minimal essential medium plus 10\% FCS and
maintained at $37^\circ$C 5\% CO2.
Cells were seeded onto 35-mm glass coverslip-based dishes (IWAKI,
Japan) and cultured for 20 h before imaging.
The dish was transferred to the stage of a Zeiss Axiovert 200
microscope equipped with an XL incubator
(maintained at $37^\circ$C, 5\% CO2, in humid conditions). Fluorescence
images were obtained using a Fluar x20,
0.75 numerical aperture (Zeiss) dry objective. Stimulus (5 $\mu$M
forskolin and 0.5 $\mu$M BayK-8644) to induce an increase in prolactin
gene expression was
added directly to the dish for 6 h, followed by treatment with 10 $\mu
$g/ml cycloheximide (for protein degradation rate)
or 3 $\mu$g/ml actinomycin D (for mRNA degradation rate) and imaged for
at least a further 15 h.
%
\subsection{The linear noise approximation}\label{appLNA}
The LNA approximates transitions densities by a Gaussian distribution
with an appropriately-defined covariance matrix.
It is usually derived as an approximation to the master equation by van
Kampen's system-size expansion [Van~Kampen
(\citeyear{VanKampen1976,VanKampen1997})].
However, here we give a simplified derivation of the LNA by \citet
{Wallace} [see also \citet{Wilkinson}] which is less general than \citet
{Kurtz1971} but more intuitive, as it treats births and deaths
separately. Start by re-expressing the rates explicitly as functions of
the system size $\Omega$,
\[
\Omega w_j \biggl(\frac{X(t)}{\Omega} \biggr).
\]
Let $w_j^+$ and $w_j^-$ denote reaction rates associated with birth and
death, respectively, and let $\eps_j(t) \sim N(0,1)$.
Then equation (\ref{Lang}) can be expressed as
\begin{eqnarray*}
X(t+\tau) & = & X(t)+ \tau\Omega\sum_{j=1}^{m}
\biggl[ w_j^+ \biggl(\frac{X(t)}{\Omega} \biggr) -w_j^-
\biggl(\frac{X(t)}{\Omega} \biggr) \biggr]v_j
\\
& &{} + \sqrt{ \tau\Omega} \sum_{j=1}^{m}
\sqrt{ w_j^+ \biggl(\frac{X(t)}{\Omega} \biggr)
+w_j^- \biggl(\frac
{X(t)}{\Omega} \biggr) } v_j \eps_j(t)
\end{eqnarray*}
or, equivalently,
%
%
\begin{eqnarray}\label{last1}
X(t+\tau) & = & X(t)+ \tau\Omega\sum_{j=1}^{m}
A_j \biggl(\frac{X(t)}{\Omega} \biggr) v_j
\nonumber\\[-8pt]\\[-8pt]
& &{} + \sqrt{ \tau\Omega} \sum_{j=1}^{m}
\sqrt{ B_j \biggl(\frac
{X(t)}{\Omega} \biggr) }v_j
\eps_j(t),
\nonumber
\end{eqnarray}
where
\begin{eqnarray*}
A_j \biggl(\frac{X(t)}{\Omega} \biggr) &=& w_j^+ \biggl(
\frac
{X(t)}{\Omega} \biggr) -w_j^- \biggl(\frac{X(t)}{\Omega} \biggr),
\\
B_j \biggl(\frac{X(t)}{\Omega} \biggr) &=& w_j^+ \biggl(
\frac
{X(t)}{\Omega} \biggr) +w_j^- \biggl(\frac{X(t)}{\Omega} \biggr).
\end{eqnarray*}
We now make the Ansatz that $X(t)$ can be decomposed into a
deterministic solution with a stochastic perturbation
the variance of which scales with $\sqrt{\Omega}$,
%
%
\begin{equation}
\label{no1} X(t)=\Omega\phi(t) + \sqrt{\Omega} \zeta(t),
\end{equation}
where $\phi(t)$ is the macroscopic or ODE solution for the
concentration and $\zeta(t)$ is a stochastic process.
Inserting this into (\ref{last1}) and dividing by $\Omega$ gives
%
%
\begin{eqnarray}\label{lastlang}
\phi(t+\tau) + \frac{1}{\sqrt{\Omega}} \zeta(t+\tau)& = & \phi (t) + \frac{1}{\sqrt{\Omega}}
\zeta(t)
\nonumber\\
& &{} + \tau\sum_{j=1}^{m} A_j
\biggl( \phi+ \frac{1}{\sqrt{\Omega
}} \zeta \biggr) v_j
\\
& &{} + \sqrt{\frac{\tau}{\Omega}} \sum_{j=1}^{m}
\sqrt{ B_j \biggl( \phi+ \frac{1}{\sqrt{\Omega}} \zeta \biggr)
}v_j \eps_j(t).
\nonumber
\end{eqnarray}
Applying a Taylor expansion to $A_j (\frac{X(t)}{\Omega}
)$ and
$ B_j (\frac{X(t)}{\Omega} ) $ about the deterministic
term $\phi$ gives
\[
A_j \biggl( \phi+ \frac{1}{\sqrt{\Omega}} \zeta \biggr) \approx
A_j(\phi) + \frac{1}{\sqrt{\Omega}} \zeta D_\phi
\bigl(A_j(\phi)\bigr) +\cdots
\]
and
\[
B_j \biggl( \phi+ \frac{1}{\sqrt{\Omega}} \zeta \biggr) \approx
B_j(\phi) + \frac{1}{\sqrt{\Omega}} \zeta D_\phi
\bigl(A_j(\phi)\bigr) +\cdots.
\]
Inserting these into (\ref{lastlang}) and collecting terms of order
$\Omega^0$ give
\[
\phi(t+\tau)=\phi(t) + \tau\sum_j
A_j(\phi) v_j
\]
or, expressing the sum in matrix form,
\[
\phi(t+\tau)=\phi(t) + \tau A(\phi),
\]
which translates into the macroscopic ODE model
%
%
\begin{equation}
\label{no2} \frac{d \phi(t)}{dt} = A(\phi).
\end{equation}
Next, collecting terms of order $\Omega^{-1/2}$ gives an equation for
the noise process
\[
\zeta(t+\tau)= \zeta(t) + \tau\sum_{j}
D_\phi\bigl(A_j(\phi)\bigr) v_j \zeta(t) +
\sqrt{\tau} \sum_{j} \sqrt{ B_j(\phi)}
v_j \eps_j(t).
\]
The corresponding SDE or Langevin form is
%
%
\begin{equation}
\label{no3} \d\zeta(t)= J(t) \zeta(t)\,dt + \sum_j
\sqrt{B_j(\phi)} \,dW_j(t),
\end{equation}
where $W_j(t)$ is a Wiener process, one for each population, and
$J(t)$ is the Jacobian matrix of the macroscopic equations
\[
J_{ij}(t)=\frac{\partial A_j(\phi)}{\partial\phi_i}= \frac
{\partial[w_j^+(\phi(t)) -w_j^-(\phi(t))]}{{\partial\phi_i}}.
\]
Equations (\ref{no1}), (\ref{no2}) and (\ref{no3}) together specify
the Linear Noise Approximation (LNA) derived by \citet{VanKampen1997}.
%

\subsection{Kalman filter and prediction error}\label{appKL}
For ease of notation we use $t$ instead of $t_i$ and $t \pm1$ instead
of $t_{i \pm1}$.
The state space model defined by (\ref{bigeuler}) and (\ref{meas})
has the form
%
%
\begin{eqnarray}
\label{se} X_{t+1} &=& F_t X_t +c_t
+ \eps_t,\qquad \eps_t \sim N( \Ov,\Sigma _{\eps,t} ),
\\
\label{me} Y_t &=& \kappa_t X_t +
u_t,\qquad u_t \sim N( \Ov,\Sigma_{u,t} ),
\end{eqnarray}
where the structure of the matrices $F_t, c_t,\Sigma_{\eps,t}$ in the
state equation (\ref{se}) is invoked by the LNA.
The measurement equation (\ref{me}) is general, allowing $\kappa
=\kappa_t$ and $\Sigma_{u}=\Sigma_{u,t}$
to vary over time. The Kalman filter is defined by the recursions given
in the following prediction and updating equations.
Let $\hat{X}_t$ denote the optimal linear estimator of $X_t$ based on
the information available at time $t$ and
let $P_t$ denote its variance matrix. Based on information up to time
$t-1$, we get the prediction equations
%
%
\begin{eqnarray}
\hat{X}_{t|t-1} & = & F_{t-1} \hat{X}_{t-1} +
c_{t-1},
\\
P_{t|t-1} & = & F_{t-1} P_{t-1} F_{t-1}^{T}
+\Sigma_{\eps,t-1}.
\end{eqnarray}
The updating equations when $Y_t$ becomes available are
%
%
\begin{eqnarray}
\hat{X}_t & = & \hat{X}_{t|t-1} + P_{t|t-1}
\kappa_t^T R_t^{-1}(Y_t-
\kappa_t \hat{X}_{t|t-1}),
\\
P_t & = & P_{t|t-1} - P_{t|t-1} \kappa_t^T
R_t^{-1} \kappa_t P_{t|t-1},
\end{eqnarray}
where
%
%
\begin{equation}
\label{KFcovariance} R_t = \kappa_t
P_{t|t-1} \kappa_t^T +\Sigma_{u,t}
\end{equation}
is the variance matrix of the prediction error
%
%
\begin{equation}
\label{KFerror} e_t = Y_t-
\kappa_t \hat{X}_{t|t-1}.
\end{equation}
The prediction error
in (\ref{KFerror}) and its variance matrix (\ref{KFcovariance} )
are used for the likelihood (\ref{PED}).

\subsection{Further details on specification of parameter distributions}

\subsubsection{Translation inhibition experiment}\label{apptransla}
The hierarchical parameters are assumed to have distributions
\[
\delta_2^{(i)}\sim\Gamma\bigl(\mu_{\delta_2},
\sigma^2_{\delta_2}\bigr),\qquad \tilde{\tau}_{2}^{(i)}
\sim\Gamma\bigl(\mu_{\tilde{\tau
}_P},\sigma^2_{\tilde{\tau}_P}\bigr),\qquad
\sigma_u^{2,(i)}\sim \Gamma\bigl(\mu_{\sigma_u},
\sigma^2_{\sigma_u}\bigr),
\]
where $\Gamma(\mu_{(\cdot)},\sigma^2_{(\cdot)})$ denotes a gamma
density parameterized to have mean $\mu_{(\cdot)}$ and variance
$\sigma^2_{(\cdot)}$.
The hyperparameters\vspace*{2pt} are
$\Theta_{H}=(\mu_{\delta_2},\sigma^2_{\delta_2},\mu_{\tilde{\tau
}_2},\sigma^2_{\tilde{\tau}_2},\mu_{\sigma_u},
\sigma^2_{\sigma_u})$.
For the prior distribution of $\Theta_{H}$ we assume a product of
vague exponential densities with parameter $10^4$ for each element of
$\Theta_{H}$.
We also reparameterize $\tilde{\phi}_2^{(i)}(0)=\kappa\phi
_P^{(i)}(0)$ assuming prior distributions $\tilde{\phi}_2^{(i)}(0)
\sim \operatorname{Exp}(10^4)$ and $\kappa\sim \operatorname{Exp}(10^4)$.
Let $\Theta=(\Theta_{H},\tilde{\phi}_2(0),\kappa)$, where $\tilde
{\phi}_2(0)$ denotes the vector of initial conditions for cells
$i=1,\ldots,N$, then
we wish to estimate $\Theta$ via their posterior distribution
as given in (\ref{Bayes}). We use a standard Metropolis--Hastings algorithm
[\citet{Gamerman,Chib1995}] to generate a sample from the posterior
distribution.

\subsubsection{Transcription inhibition experiment}\label{apptranscript}
To improve convergence of the estimation algorithm for the
two-dimensional model, we reparameterized
$\tilde{\tau}_1^{(i)} = \kappa\alpha^{(i)} \tau_1^{(i)}$ and
$\tilde{\alpha}^{(i)}=\kappa\alpha^{(i)}$,
as well as the initial conditions $\tilde{\phi}_1^{(i)}(0)=\kappa
\alpha^{(i)} \phi_1^{(i)}(0)$ and $\tilde{\phi}_2^{(i)}(0)=\kappa
\phi_2^{(i)}(0)$.
The hierarchical parameters are assumed to have the following distributions:
\begin{eqnarray*}
\delta_1^{(i)}&\sim&\Gamma\bigl(\mu_{\delta_1},
\sigma^2_{\delta_1}\bigr),\qquad \tilde{\tau}_{1}^{(i)}
\sim\Gamma\bigl(\mu_{\tilde{\tau}_1},\sigma
^2_{\tilde{\tau}_1}\bigr),\\
\tilde{\alpha}^{(i)}&\sim&\Gamma\bigl(\mu _{\tilde{\alpha}},
\sigma^2_{\tilde{\alpha}}\bigr),\qquad
\sigma _u^{2,(i)}
\sim\Gamma\bigl(\mu_{\sigma_u},\sigma^2_{\sigma_u}\bigr).
\end{eqnarray*}
We have used a vague prior for $\sigma_u^{2,(i)}$ rather than importing
a prior informed by the translation inhibitor experiment, as it is
possible that the setting of the experiment and the use of a camera
might have resulted
in a very different variance of the measurement error.
The vector of hyperparameters is
$\Theta_{H}=(\mu_{\delta_1},\sigma^2_{\delta_1},\mu_{\tilde{\tau
}_1},\sigma^2_{\tilde{\tau}_1},\mu_{\tilde{\alpha}},\sigma
^2_{\tilde{\alpha}}, \mu_{\sigma_u},\sigma^2_{\sigma_u})$
and the full parameter vector is
$\Theta=(\Theta_H,\tilde{\phi}_1(0),\tilde{\phi}_2(0),\kappa)$,
where $\tilde{\phi}_1(0) $ and
$\tilde{\phi}_2(0)$ denote the vectors of unknown initial conditions
for mRNA and protein, respectively.
Similar to the previous experiment, the prior for each element of
$\Theta$ was $\operatorname{Exp}(10^4)$ except we
set $\mu_{\delta_2}=0.57$ and $\sigma^2_{\delta_2}=0.004$,
importing the posterior results on protein degradation from
the translation inhibitor experiment.
\end{appendix}

\section*{Acknowledgments}

The authors wish to thank Kirsty Hey (Department of Statistics,
University of Warwick) and an anonymous referee for valuable
suggestions. MK was funded during Ph.D. by a scholarship from
University of Warwick. Hamamatsu Photonics and Carl Zeiss Limited
provided technical support. The Endocrinology Group, University of
Manchester, is supported by the Manchester Academic Health Sciences
Centre (MAHSC) and the NIHR Manchester Biomedical Research Centre.

\section*{Author contribution}

DJW and MK performed numerical estimations supervised by BF. DJW
constructed the MCMC sampler for the transcription inhibition model and
MK for the translation inhibition. BF proposed the use of hierarchical
modeling and wrote the manuscript with input from DJW, MK and DAR. CVH
devised the experimental single cell degradation approach and performed
and analyzed the experiments supervised by JRED and MRHW. DAR provided
help on the mathematical modeling.



\printaddresses

\end{document}